\documentclass[sigconf,nonacm]{acmart}

\usepackage{acronym}
\usepackage[ruled,vlined,linesnumbered]{algorithm2e}
\usepackage[english]{babel}
\usepackage{color}
\usepackage[inline]{enumitem}
\usepackage{float}
\usepackage{fontawesome}
\usepackage{makecell}
\usepackage{multirow}
\usepackage{subcaption}
\usepackage[table]{xcolor}
\usepackage{xcolor}

\usepackage{amssymb}

\SetKwComment{rtri}{$\triangleright$ }{}
\SetKwComment{rbtri}{$\blacktriangleright$ }{}

\newcommand{\heading}[1]{\par\noindent\textbf{#1.}}

\AtBeginDocument{
  \providecommand\BibTeX{{
    \normalfont B\kern-0.5em{\scshape i\kern-0.25em b}\kern-0.8em\TeX}}}

\makeatletter
\g@addto@macro\normalsize{
  \abovedisplayskip 3pt plus1pt
  \belowdisplayskip 3pt plus1pt
  \abovedisplayshortskip  0pt plus1pt
  \belowdisplayshortskip  0pt plus1pt
}
\makeatother

\acrodef{CV}{computer vision}
\acrodef{IR}{information retrieval}
\acrodef{LLM}{large language model}
\acrodef{MDP}{Markov decision process}
\acrodef{NLP}{natural language processing}
\acrodef{NRM}{neural ranking model}
\acrodef{RL}{reinforcement learning}
\acrodef{RL-MARA}{Multi-grAnular Ranking Attack}
\acrodef{MoE}{mixture-of-experts}

\copyrightyear{2026}
\acmYear{2026}
\setcopyright{cc}
\setcctype{by}
\acmISBN{}
\acmDOI{}
\begin{document}

\title[CLEAR: Structured Specifications for Controllable Long-Horizon ML Research Engineering]{CLEAR: Structured Specifications for Controllable Long-Horizon ML Research Engineering}

\begin{abstract}
Large language model (LLM) agents are increasingly capable of automating machine learning (ML) research engineering, including code development, experiment execution, and research reproduction. 
However, existing systems typically couple research interpretation directly with code generation and execution, making them vulnerable to cascading errors and costly end-to-end repair. 
We identify a fundamental cause of this limitation: \emph{research artifacts are not executable specifications}. 
They describe models, data, algorithms, and evaluation protocols through heterogeneous and often implicit representations, leaving critical implementation decisions to be resolved during generation.

We propose CLEAR, a specification-centered framework for controllable long-horizon ML research engineering. CLEAR introduces an explicit intermediate layer that transforms heterogeneous research artifacts into structured and human verifiable execution specifications before executable generation. These specifications provide persistent states for decomposing workflows into localized implementation units, enabling specification-conditioned generation and repair rather than repeated end-to-end regeneration. Building on this representation, CLEAR combines a strong planning model with a dynamically adjusted number of low-cost coding workers according to task decomposition complexity, and employs a lifecycle-oriented harness to monitor workflow states, localize failures, and perform cost-aware intervention.
Experiments on PaperBench and MLE-Bench Lite demonstrate that CLEAR achieves a favorable performance--efficiency trade-off.
The code is available at \textcolor{blue}{\url{https://anonymous.4open.science/r/CLEAR-VSS}.}
\end{abstract}

\ccsdesc[500]{Computing methodologies~Planning and scheduling}
\ccsdesc[500]{Computing methodologies~Multi-agent systems}

\keywords{Structured execution specifications, Long-horizon automation}

\author{Xin Liu}
\orcid{0009-0000-8548-0936}
\affiliation{
 \institution{State Key Laboratory of AI Safety}
 \institution{Institute of Computing Technology, Chinese Academy of Sciences}
 \institution{University of Chinese Academy of Sciences}
 \city{Beijing}
 \country{China}
}
\email{liuxin25s@ict.ac.cn}

\author{Hongru Song}
\orcid{0009-0000-8443-9499}
\affiliation{
 \institution{State Key Laboratory of AI Safety}
 \institution{Institute of Computing Technology, Chinese Academy of Sciences}
 \institution{University of Chinese Academy of Sciences}
 \city{Beijing}
 \country{China}
}
\email{songhongru24s@ict.ac.cn}

\author{Ruqing Zhang}
\authornote{Ruqing Zhang is the corresponding author.}
\orcid{0000-0003-4294-2541}
\affiliation{
 \institution{State Key Laboratory of AI Safety}
 \institution{Institute of Computing Technology, Chinese Academy of Sciences}
 \institution{University of Chinese Academy of Sciences}
 \city{Beijing}
 \country{China}
}
\email{zhangruqing@ict.ac.cn}

\author{Jiafeng Guo}
\orcid{0000-0002-9509-8674}
\affiliation{
 \institution{State Key Laboratory of AI Safety}
 \institution{Institute of Computing Technology, Chinese Academy of Sciences}
 \institution{University of Chinese Academy of Sciences}
 \city{Beijing}
 \country{China}
}
\email{guojiafeng@ict.ac.cn}

\author{Xueqi Cheng}
\orcid{0000-0002-5201-8195}
\affiliation{
 \institution{State Key Laboratory of AI Safety}
 \institution{Institute of Computing Technology, Chinese Academy of Sciences}
 \institution{University of Chinese Academy of Sciences}
 \city{Beijing}
 \country{China}
}
\email{cxq@ict.ac.cn}

\renewcommand{\shortauthors}{Xin Liu et al.}

\maketitle

\section{Introduction}
Large language model (LLM) agents are increasingly being used to automate machine learning (ML) research workflows, ranging from code development and experiment execution to research reproduction and scientific discovery~\citep{lu2025automated,huang2024mlagentbench,zhang2025darwin,wang2024autosurvey}. 
Recent systems demonstrate that LLMs can perform substantial portions of ML research engineering with limited human intervention~\citep{starace2025paperbench,mohammadshahi2024leeroo,jiang2025aide,zhang2026aibuildai}. 
These advances suggest a shift from automating individual research tasks toward automating long-horizon workflows that transform research artifacts into executable ML systems.

However, ML research engineering remains substantially more difficult than isolated code generation. 
A research artifact typically specifies a system only implicitly through a mixture of natural-language descriptions, equations, pseudo-code, implementation details, datasets, and evaluation protocols. 
An agent must first infer the intended computational structure, translate it into executable components, coordinate their dependencies, and repeatedly diagnose and repair failures during execution. 
Consequently, the quality of the final system depends not only on code generation, but also on whether the agent can maintain a consistent representation of what should be implemented throughout this long-horizon process.

\begin{figure*}[t]
  \centering
  \includegraphics[width=\textwidth]{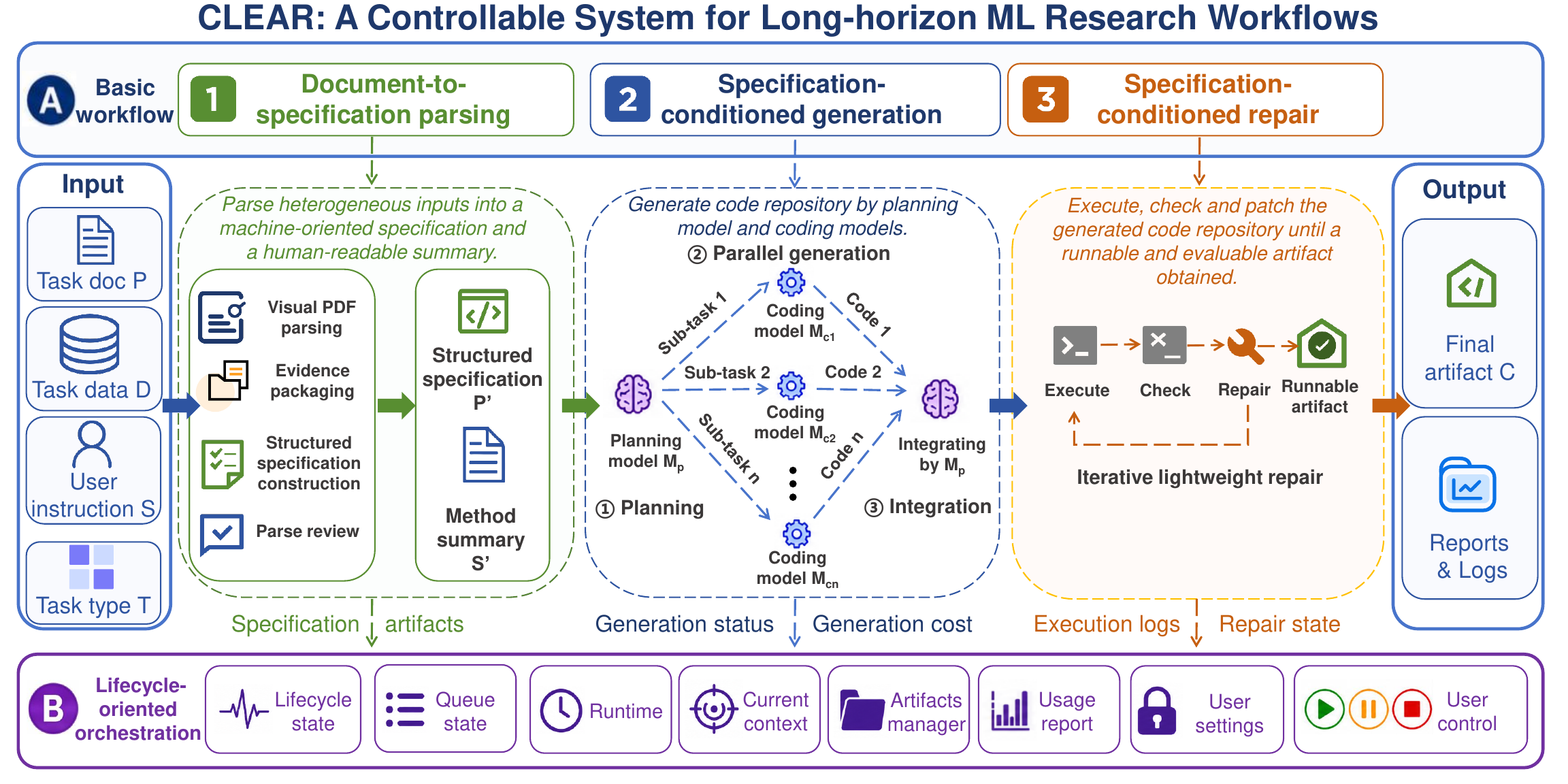}
  \caption{Overview of CLEAR. Structured execution specifications bridge heterogeneous research artifacts and executable implementations, enabling specification-conditioned generation, specification-conditioned repair, and lifecycle-oriented orchestration for long-horizon ML research engineering.}
  \label{fig:workflow}
  \vspace{2mm}
\end{figure*}

\heading{The Missing Layer: Structured Execution Specifications}
Existing systems largely operate without such an explicit intermediate representation. They directly transform heterogeneous research artifacts into plans, code, and execution traces within a tightly coupled agent loop. 
This design creates a fundamental problem: \emph{research artifacts are not executable specifications}. 
Their contents often leave implicit implementation assumptions, while different parts of a paper may specify model architecture, data processing, training procedures, and evaluation criteria at different levels of abstraction. When these ambiguities are resolved implicitly during generation, errors introduced during early interpretation can become embedded in downstream code and execution states. Once failures occur, existing systems commonly recover by regenerating large portions of the workflow, further increasing computational cost and making error recovery unstable.

We identify three consequences of this missing intermediate representation. 
First, \emph{representation ambiguity} makes it difficult to distinguish what the research artifact explicitly specifies from what the agent infers during implementation. 
Second, \emph{execution coupling} entangles interpretation, planning, coding, and debugging, allowing early mistakes to propagate across dependent components. 
Third, \emph{global repair} forces the system to repeatedly regenerate already-correct parts of the workflow when only a localized component has failed. 
These issues reveal a common underlying limitation: existing agents lack a persistent, inspectable state that separates \emph{what should be implemented} from \emph{how it is implemented and repaired}.

\heading{CLEAR: Specification-Centered and Controllable Automation}
To address this limitation, we propose \textbf{CLEAR}, a \underline{C}ontrollable \underline{L}ong-horizon \underline{E}ngineering system for \underline{A}utomating ML \underline{R}esearch workflows. 
CLEAR introduces a structured specification layer between research artifacts and executable implementations. 
Instead of using papers and other artifacts as direct prompts for code generation, CLEAR first converts them into structured and human verifiable execution specifications that explicitly describe implementation units, dependencies, interfaces, execution requirements, and evaluation procedures. 
These specifications serve as intermediate states that can be referenced at any time throughout the entire workflow, allowing downstream generation and repair to operate on localized, well-defined objectives rather than repeatedly interpreting the original artifacts.

This intermediate representation also enables two forms of controllability that are difficult to achieve in end-to-end agent loops. 
First, CLEAR performs \emph{specification-conditioned model routing}: large models are used for global interpretation, planning, and integration, while smaller models handle specification-bounded code generation and localized repair. Because each generation task is constrained by an explicit specification, the number of smaller models used in generation can be allocated according to the task complexity.
Second, CLEAR introduces a \emph{lifecycle-oriented execution harness} that treats research engineering as a sequence of observable workflow states rather than an opaque generation loop. The harness tracks the specification, generation, execution, failure, and repair states, enabling targeted intervention, runtime observability, and cost-aware control.

As illustrated in Figure~\ref{fig:workflow}, CLEAR therefore follows a specification-centered workflow consisting of three stages. \emph{(1) Document-to-specification parsing} converts heterogeneous research artifacts into structured execution specifications that expose implementation requirements and dependencies. \emph{(2) Specification-conditioned generation and repair} decomposes the workflow into specification-bounded units and assigns number of smaller coding models according to the complexity of each unit, enabling localized generation and recovery. \emph{(3) Lifecycle-oriented orchestration} maintains the state of the workflow during execution, monitors failures and costs, and triggers targeted repair instead of indiscriminate end-to-end regeneration. Together, these components transform ML research engineering from a tightly coupled generation loop into a controllable sequence of specification-grounded state transitions.

The central design principle of CLEAR is therefore to introduce an explicit intermediate representation between research understanding and executable realization. This separation provides a common foundation for both \emph{reliability} and \emph{efficiency}: specifications reduce ambiguity and localize failures, while specification-bounded tasks enable selective model amount allocation and targeted repair. Rather than increasing model size to compensate for increasingly complex workflows, CLEAR uses structured intermediate states to control where model capacity and computation are needed.

Our contributions are summarized as follows:
\begin{enumerate}[leftmargin=*]

\item We identify the lack of an explicit intermediate representation as a fundamental bottleneck in long-horizon ML research engineering. We formulate structured execution specifications as a persistent state that separates research interpretation from executable realization and enables localized downstream control.

\item We propose CLEAR, a specification-centered framework that converts heterogeneous research artifacts into human verifiable execution specifications and uses them to support specification-conditioned generation, and specification-conditioned repair.

\item We develop a lifecycle-oriented execution harness that provides fine-grained control and observability over specification, generation, execution, failure, and repair states, enabling cost-aware intervention throughout long-horizon research workflows.

\item Extensive experiments on PaperBench and MLE-Bench Lite show that CLEAR achieves a favorable performance--efficiency trade-off. On PaperBench, CLEAR obtains a score of 29.87 at an average cost of \$7.91 per paper, reducing cost by 35.16\% relative to \emph{AiScientist} while sacrificing only 3.86 points in performance. On MLE-Bench Lite, CLEAR achieves 77.27\% Any Medal while substantially reducing repair iterations on representative tasks.

\end{enumerate}

\section{Task Formulation}
\label{Task_Formulation}

\heading{Problem Definition}
We formulate ML research engineering as the problem of transforming heterogeneous research artifacts into executable and evaluable computational artifacts through a sequence of interpretation, implementation, and execution decisions. Unlike conventional code generation, the input to ML research engineering may contain natural-language descriptions, equations, pseudo-code, implementation details, datasets, and evaluation protocols, which jointly specify the intended system at different levels of abstraction.

Formally, given a task type $T$, a research artifact $P$, optional task data $D$, and user instructions $S$, an agent system $\mathcal{A}$ produces an executable artifact $C$:
\begin{equation}
    C = \mathcal{A}(T, P, D, S).
\end{equation}
The generated artifact $C$ may include a runnable code repository, experiment pipeline, trained model, evaluation results, or valid competition submission, depending on the task type.

A valid solution must satisfy two requirements: (i) \emph{executability}, meaning that the artifact can be successfully executed or submitted under the task environment; and (ii) \emph{task performance}, measured by a task-specific evaluation function. We therefore formulate ML research engineering as the following constrained optimization problem:
\begin{equation}
\begin{aligned}
    C^\star
    &= \arg\max_{C \in \mathcal{C}_T} Q_T(C), \\
    \text{s.t.}\quad
    &\mathrm{Exec}_T(C)=1,
\end{aligned}
\end{equation}
where $\mathcal{C}_T$ denotes the artifact space for task type $T$, $Q_T(\cdot)$ is the task-specific evaluation metric, and $\mathrm{Exec}_T(C)$ indicates whether $C$ can be executed or submitted successfully.

\heading{Task Instantiations} In this work, we consider two representative task types:

\begin{enumerate}[leftmargin=*]

\item \emph{Replicate AI Research (RAR)}, where the input is an AI research paper $P_r$ and the goal is to construct a runnable reproduction artifact:
\begin{equation}
    C_r = \mathcal{A}(\mathrm{RAR}, P_r, \varnothing, S_r).
\end{equation}

\item \emph{Machine Learning Engineering (MLE)}, where the input consists of a task description $P_m$ and task data $D_m$, and the goal is to construct a runnable ML pipeline or a valid competition submission:
\begin{equation}
    C_m = \mathcal{A}(\mathrm{MLE}, P_m, D_m, S_m).
\end{equation}

\end{enumerate}

Although RAR and MLE differ in their input artifacts and evaluation protocols, they share the same underlying structure: both require an agent to interpret heterogeneous task information, translate it into implementation decisions, and iteratively construct and execute a multi-step computational workflow. This shared structure makes them suitable settings for studying long-horizon ML research engineering.

To capture this common structure, we distinguish the \emph{research artifact} provided as input from the \emph{execution specification} required for implementation. A research artifact describes the intended research or task objectives, often implicitly and across heterogeneous formats. In contrast, an execution specification provides an explicit representation of what needs to be implemented and how its components are related. In CLEAR, this specification consists of two complementary views: a machine-oriented specification $P'$ and a human-readable method summary $S'$. We therefore formulate the workflow as
\begin{equation}
(P,D,S)
\xrightarrow{\mathcal{A}_{\mathrm{spec}}}
(P',S')
\xrightarrow{\mathcal{A}_{\mathrm{exec}}}
C,
\end{equation}
where $(P',S')$ serves as the intermediate representation between research understanding and executable realization.

This formulation separates \emph{what should be implemented}, represented by $(P',S')$, from \emph{how it is realized}, represented by $\mathcal{A}_{\mathrm{exec}}$.

\section{Our Method}
\label{Our_Method}

Given the formulation above, CLEAR operationalizes $(P',S')$ as an intermediate state throughout long-horizon ML research engineering. The system follows a \emph{specification-first} design principle: research artifacts are first transformed into structured execution specifications, after which planning, generation, execution, and repair are performed with respect to these specifications. This decomposition replaces the repeated interpretation of raw research artifacts with specification-bounded implementation and repair decisions. As shown in Figure~\ref{fig:workflow}, CLEAR consists of three stages:

\begin{enumerate}[leftmargin=*]

\item \emph{Document-to-Specification Parsing}, where the system receives the task type $T$, research artifact $P$, optional task data $D$, and optional user instruction $S$, and transforms these heterogeneous inputs into the structured execution specification $(P',S')$. The machine-oriented specification $P'$ captures implementation requirements, component dependencies, interfaces, execution constraints, and evaluation procedures, while the human-readable summary $S'$ provides a concise abstraction of the intended research method for downstream planning and coordination.

\item \emph{Specification-Conditioned Generation}, where a strong planning model uses $(P',S')$ to decompose the workflow into localized, specification-bounded subtasks and coordinates lower-cost coding models to generate the initial repository $C_0$. The explicit specification allows the number of lower-cost coding workers to be adjusted according to task complexity and enables independent subtasks to be generated in parallel while preserving their specified dependencies and interfaces.

\item \emph{Lifecycle-Controlled Execution and Repair}, where the generated repository $C_0$ is executed and evaluated under the task environment. When failures occur, CLEAR identifies affected subtasks and performs targeted, specification-conditioned iterative repair rather than regenerating the entire workflow. The process iterates through execution, diagnosis, and repair until a runnable and evaluable artifact $C$ is obtained.

\end{enumerate}

A \emph{visual lifecycle harness} serves as the control layer across all three stages. Rather than treating LLM agents as a single black-box generator, the harness explicitly controls when and how agents parse, plan, generate, execute, and repair. It exposes intermediate specifications and execution artifacts for inspection and user intervention, while tracking runtime, token usage, and estimated cost. This lifecycle-oriented control makes the workflow observable and steerable, while the specification-centered decomposition enables CLEAR to localize failures, allocate the number of lower-cost coding model selectively, and reduce unnecessary regeneration throughout long-horizon ML research engineering.

\begin{figure}[t]
\centering
\includegraphics[width=\columnwidth]{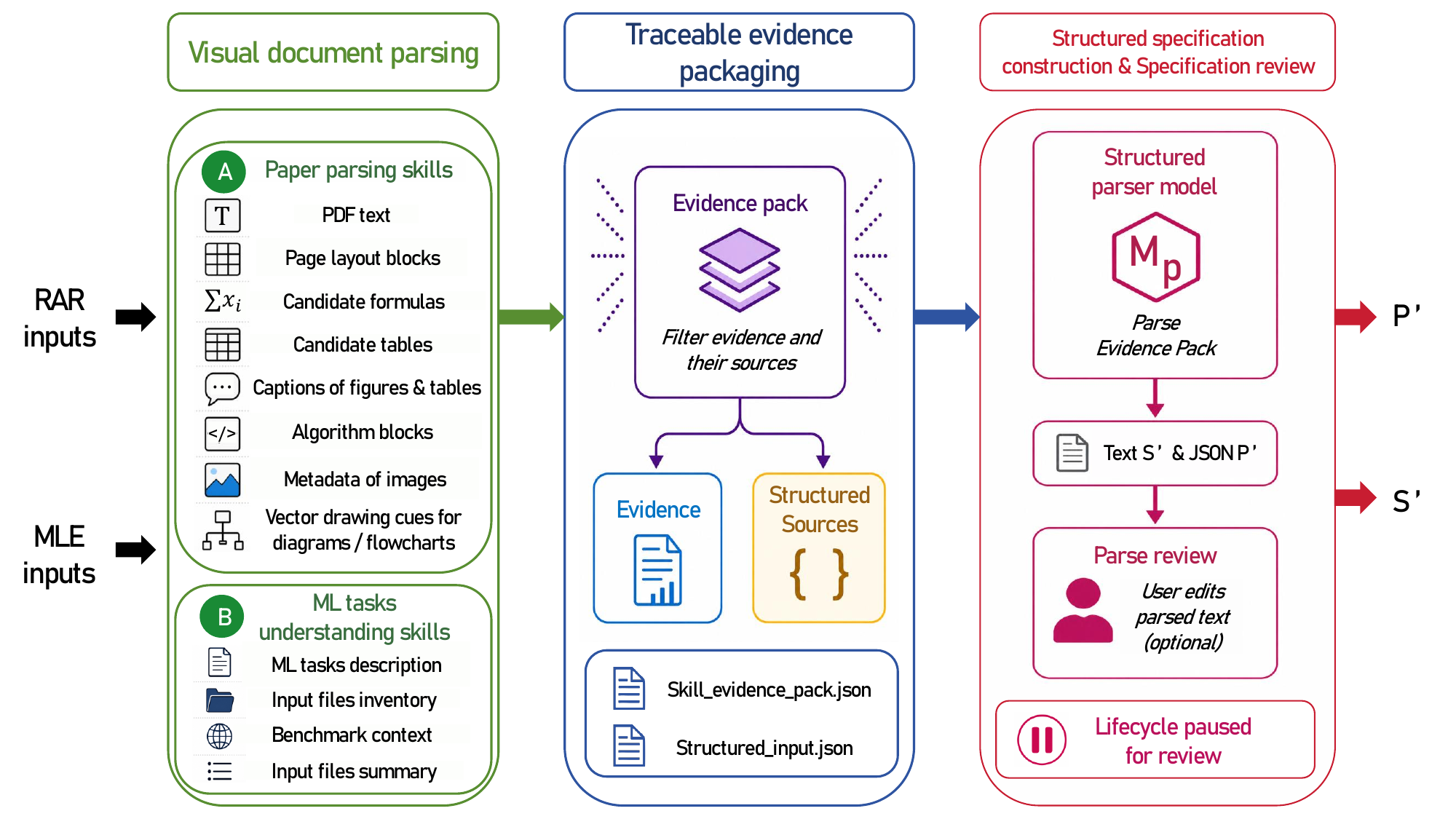}
\caption{Document-to-specification parsing in CLEAR. Heterogeneous research artifacts are parsed into traceable evidence units and synthesized into complementary machine-oriented and human-readable specification views $(P',S')$. An optional review step provides a control point for resolving ambiguous requirements before downstream code generation.}
\label{fig:skills}
\end{figure}

\subsection{Document-to-Specification Parsing}
\label{sec:specification_parsing}

The first stage of CLEAR constructs the execution specification that serves as the intermediate state for downstream implementation. Given a task type $T$, research artifact $P$, optional task data $D$, and user instruction $S$, CLEAR transforms these heterogeneous inputs into two complementary views: a machine-oriented structured specification $P'$ and a human-readable method summary $S'$. Rather than providing another free-form description of the input, $(P',S')$ explicitly captures what needs to be implemented and how the resulting system should be executed and evaluated.

We formalize the parsing process as
\begin{equation}
    (P', S') = g(T, P, D, S),
\end{equation}
where $g$ denotes the document-to-specification process. The structured specification $P'$ records implementation-relevant requirements, including algorithms, inputs and outputs, dependencies, parameters, interfaces, evaluation metrics, and submission constraints. The summary $S'$ provides a concise and human-readable view of the inferred research method and serves as an inspection interface for users. Both views are derived from the same underlying evidence and jointly represent the execution specification used throughout the subsequent workflow.

As illustrated in Figure~\ref{fig:skills}, CLEAR constructs $(P',S')$ through three processing steps, followed by an optional review step:

\begin{enumerate}[leftmargin=*]

\item \emph{Visual document parsing}, which extracts text, page structure, formulas, tables, figures, captions, and algorithmic descriptions from the input research artifact, preserving both semantic content and document structure.

\item \emph{Traceable evidence packaging}, which identifies implementation-relevant evidence and organizes it into traceable evidence units. Each unit retains its source context, allowing specification entries to remain grounded in the original research artifact rather than being generated as unsupported summaries.

\item \emph{Structured specification construction}, where the planning model $M_p$ synthesizes the evidence units into the structured specification $P'$ and the summary of the corresponding method $S'$. The resulting representation explicitly organizes the implementation requirements, component relationships, execution constraints, and evaluation procedures.

\item \emph{Specification review}, where users can inspect $S'$ and, when necessary, correct or refine the inferred understanding before code generation. This optional intervention provides an explicit control point for resolving ambiguous or incorrectly interpreted requirements before they propagate to downstream implementation.

\end{enumerate}

The resulting $(P',S')$ therefore transforms implicit and heterogeneous research understanding into an explicit, traceable, and reviewable execution state. This separation allows downstream planning, generation, and repair to operate on specification-bounded units rather than repeatedly interpreting the original research artifacts, reducing ambiguity and limiting the propagation of early-stage interpretation errors.

\subsection{Specification-Conditioned Generation}
Given the execution specification $(P',S')$, CLEAR generates the initial repository $C_0$ through specification-bounded decomposition and heterogeneous model collaboration. The key idea is to separate \emph{global reasoning} from \emph{localized implementation}: the stronger planning model $M_p$ handles algorithm interpretation, project-level decomposition, and cross-component integration, while lower-cost coding models $M_c$ implement individual subtasks whose requirements and interfaces are explicitly specified by $(P',S')$. 

The generation process consists of three steps: 
\begin{enumerate}[leftmargin=*]

\item \emph{Specification-guided planning.}
The planning model $M_p$ jointly reads $P'$ and $S'$ and decomposes the implementation into $n$ specification-bounded subtasks $I_{1:n}=\{I_1,\ldots,I_n\}$. Each subtask specifies its implementation requirements, relevant files, dependencies, interfaces, and expected outputs, so that localized workers can operate without repeatedly interpreting the original research artifact.

\item \emph{Parallel local generation.}
For each subtask $I_j$, a lower-cost coding worker $M_c$ generates the required code under the specified file paths and interfaces. Independent subtasks can be processed in parallel, allowing the number of low-cost coding workers to be dynamically allocated for local implementation without repeatedly invoking the stronger planning model.

\item \emph{Specification-aware integration.}
The planning model $M_p$ collects the generated components and verifies their  dependencies, interfaces, file structure, and cross-component consistency against $(P',S')$. It resolves conflicts when necessary and integrates the validated components into the initial repository $C_0$.

\end{enumerate} 

We formalize the generation stage as
\begin{equation}
    C_0 = G(P', S'; M_p, M_c, n),
\end{equation}
where $G$ denotes the specification-conditioned generation process, $M_p$ performs global planning and integration, $M_c$ performs localized implementation. Here, $n$ controls the degree of parallel decomposition: more complex workflows with more independent implementation units activate more coding workers, whereas simpler workflows require fewer workers.

This design turns long-horizon implementation into a collection of specification-bounded local decisions while preserving global coordination. The stronger model is therefore reserved for tasks that require project-level reasoning, whereas lower-cost models handle routine and parallelizable coding work, improving the performance--cost trade-off of generation.

\subsection{Lifecycle-Controlled Execution and Repair}
The generated repository $C_0$ is not assumed to be correct. CLEAR therefore  treats execution and repair as a controlled lifecycle rather than a single post-generation verification step. The lifecycle harness repeatedly executes the repository, observes execution feedback, localizes failures, and invokes targeted repair until a runnable and evaluable artifact $C$ is obtained.
Specifically, the process consists of three steps:

\begin{enumerate}[leftmargin=*]

\item \emph{Execution and diagnosis.}
CLEAR executes $C_0$ in the task environment and collects runtime output, error messages, test results, and task-specific evaluation feedback. The harness uses these observations to identify the affected implementation components and determine whether further repair is required.

\item \emph{Specification-conditioned repair.}
When a failure is detected, the coding model $M_c$ receives the relevant execution feedback together with the corresponding specification-bounded subtask and modifies only the affected files or components. Unlike full repository regeneration, this localized repair preserves already validated components and avoids unnecessary model computation.

\item \emph{Task-specific validation.}
The repaired repository is re-executed and evaluated under the task requirements. For MLE tasks, this includes evaluation on the provided task data $D$ and submission-specific checks, such as competition submission validation. The lifecycle continues until the artifact satisfies the required execution and evaluation conditions or reaches the predefined intervention limit.

\end{enumerate}

We formalize the execution and repair process as
\begin{equation}
    C = E(T, D, C_0; M_c),
\end{equation}
where $E$ denotes the lifecycle-controlled execution and repair process. 
The resulting artifact $C$ is therefore produced through an iterative loop:
\begin{equation}
    C_0
    \xrightarrow{\mathrm{execute}}
    O_1
    \xrightarrow{\mathrm{diagnose}}
    I_j
    \xrightarrow{\mathrm{repair}}
    C_1
    \xrightarrow{\mathrm{execute}}
    \cdots
    \xrightarrow{}
    C,
\end{equation}
where $O_t$ denotes the execution and evaluation feedback at iteration $t$.
By localizing repair to specification-bounded components and using execution feedback to guide subsequent actions, CLEAR avoids unnecessary full-workflow regeneration and preserves successful intermediate components. The lifecycle harness further provides runtime observability and control over the generation, execution, and repair states, enabling cost-aware intervention throughout the long-horizon workflow.

\subsection{Lifecycle-Oriented Orchestration}
\label{sec:lifecycle}

Long-horizon ML research engineering involves multiple dependent stages, from document parsing and user review to generation, execution, and repair. Failures or costly decisions can occur at any stage, making a monolithic agent difficult to monitor, interrupt, and recover. CLEAR therefore treats the workflow as a sequence of observable and controllable lifecycle states, managed by a visual execution harness.

\heading{Interruptible and recoverable control}
CLEAR maintains task states, inputs, and intermediate artifacts in a persistent priority queue. This enables concurrent task scheduling as well as recovery from service interruptions without restarting completed stages. Users can pause, stop, or resume tasks through the frontend, with control actions applied at safe lifecycle boundaries. The harness also records failures, timeouts, interruptions, and recovery states for subsequent diagnosis.

\heading{Human-in-the-loop checkpoints}
CLEAR introduces an explicit checkpoint after document-to-specification parsing, where the generated method summary $S'$ is presented to the user before code generation begins. Users can inspect and revise the parsed understanding, correcting missing assumptions, parameters, or evaluation requirements before they propagate to downstream implementation. The workflow resumes only after the user confirms or updates $S'$.

\heading{Observable execution}
CLEAR decomposes the workflow into 15 operational stages that span parsing, planning, generation, execution, validation, evaluation, and output archiving (as shown in the document of our anonymous repository \textcolor{blue}{\url{https://anonymous.4open.science/r/CLEAR-VSS/CLEAR_Running_Stages.pdf}}). The harness reports the current state of the lifecycle and exposes relevant intermediate artifacts, execution logs, and repair states through the frontend. In addition, it tracks runtime, token consumption, and estimated model cost, providing both users and the system with visibility into execution progress and resource usage.

In general, the lifecycle harness transforms CLEAR from a black-box code generator into an observable and steerable execution system. The persistent state enables recovery, checkpoints allow targeted human intervention, and stage-level observability enables cost-aware monitoring and diagnosis throughout the long-horizon workflow.

\begin{table*}[t]
\centering
\caption{Main results on the full PaperBench evaluation. Bold and underlined denote the best and second-best results, respectively. The blue shaded column highlights CLEAR (Ours).}
\label{tab:paperbench}

\setlength{\tabcolsep}{10pt}
\renewcommand{\arraystretch}{1.12}

\definecolor{ourcol}{RGB}{238,244,255}
\definecolor{headergray}{RGB}{245,245,245}

\begin{tabular}{lccccccc}
\toprule
& \multicolumn{3}{c}{\textbf{Gemini}} 
& \multicolumn{3}{c}{\textbf{GLM}}
& \textbf{Opus \& Sonnet} \\
\cmidrule(lr){2-4}
\cmidrule(lr){5-7}

\textbf{Task}
& \textbf{Basic}
& \textbf{Iter}
& \textbf{AiScientist}
& \textbf{Basic}
& \textbf{Iter}
& \textbf{AiScientist}
& \cellcolor{ourcol}\textbf{CLEAR} \\
\midrule

adaptive-pruning
& 24.53 & 3.05 & 27.25
& \underline{30.82} & 11.93 & \textbf{33.26}
& \cellcolor{ourcol}21.17 \\

all-in-one
& 20.86 & 45.13 & 46.29
& 33.78 & 44.43 & \textbf{49.47}
& \cellcolor{ourcol}\underline{48.20} \\

bam
& 48.46 & 45.04 & \underline{56.59}
& 51.45 & 47.91 & \textbf{61.11}
& \cellcolor{ourcol}52.33 \\

bbox
& 15.43 & 8.30 & \textbf{33.79}
& 23.55 & 19.28 & \underline{30.02}
& \cellcolor{ourcol}28.23 \\

bridging-data-gaps
& 12.59 & 12.44 & 23.09
& 9.80 & 12.50 & \textbf{26.46}
& \cellcolor{ourcol}\underline{26.16} \\

fre
& 21.67 & 23.89 & \textbf{35.21}
& 21.60 & 16.67 & \underline{28.98}
& \cellcolor{ourcol}28.71 \\

ftrl
& 5.87 & 4.15 & \textbf{10.11}
& 3.71 & 6.70 & 8.34
& \cellcolor{ourcol}\underline{8.82} \\

lbcs
& 17.75 & 15.26 & \underline{27.90}
& 20.68 & 22.74 & \textbf{30.10}
& \cellcolor{ourcol}26.24 \\

lca-on-the-line
& 12.97 & 18.30 & \textbf{30.23}
& 22.55 & 26.15 & \underline{28.53}
& \cellcolor{ourcol}24.64 \\

mechanistic-understanding
& 14.86 & 21.89 & 29.95
& 32.49 & 34.96 & \textbf{40.55}
& \cellcolor{ourcol}\underline{35.69} \\

pinn
& 26.63 & 30.81 & 49.92
& 22.18 & 25.77 & \textbf{58.76}
& \cellcolor{ourcol}\underline{49.97} \\

rice
& 10.43 & 8.88 & \textbf{10.87}
& 6.56 & 0.27 & 10.18
& \cellcolor{ourcol}\underline{10.58} \\

robust-clip
& 15.45 & 10.43 & 18.28
& 22.43 & \underline{27.56} & \textbf{28.66}
& \cellcolor{ourcol}19.82 \\

sample-specific-masks
& 25.39 & 33.34 & 36.77
& 36.93 & \underline{41.26} & \textbf{44.13}
& \cellcolor{ourcol}37.35 \\

sapg
& 11.45 & 12.65 & 19.85
& 6.99 & 4.95 & \textbf{31.69}
& \cellcolor{ourcol}\underline{24.19} \\

sequential-neural
& 53.51 & \underline{60.24} & \textbf{64.94}
& 27.20 & 35.53 & 49.32
& \cellcolor{ourcol}49.85 \\

stay-on-topic
& 8.37 & 13.69 & \textbf{20.13}
& 3.69 & 8.81 & 14.81
& \cellcolor{ourcol}\underline{16.26} \\

stochastic-interpolants
& 17.04 & 17.37 & 18.81
& 32.18 & 28.06 & \textbf{42.10}
& \cellcolor{ourcol}\underline{39.77} \\

test-time-model-adaptation
& 15.27 & 18.13 & \textbf{32.45}
& 17.81 & 21.19 & 27.33
& \cellcolor{ourcol}\underline{28.46} \\

what-will-my-model-forget
& 6.61 & 8.99 & 17.87
& \underline{25.14} & 10.75 & \textbf{30.82}
& \cellcolor{ourcol}20.98 \\

\midrule

\rowcolor{headergray}
\textbf{Average Score}
& 19.26 & 20.60 & \underline{30.52}
& 22.58 & 22.37 & \textbf{33.73}
& \cellcolor{ourcol}\textbf{29.87} \\

\rowcolor{headergray}
\textbf{Average Cost}
& \underline{\$6.25} & \$27.44 & \$15.67
& \textbf{\$4.90} & \$54.90 & \$12.20
& \cellcolor{ourcol}\$7.91 \\
\bottomrule
\end{tabular}
\vspace{2mm}
\end{table*}

\section{Experiment Settings}

\heading{Datasets}
We consider two ML research engineering tasks, RAR and MLE (Section \ref{Task_Formulation}), and evaluate CLEAR using one dataset per task.  
\begin{itemize}[leftmargin=*]
\item For RAR tasks, PaperBench \citep{starace2025paperbench} evaluates whether agents can understand 20 top-tier ML papers from scratch, implement their codebases, and successfully run experiments.

\item For MLE tasks, MLE-Bench \citep{chan2025mle} evaluates agents' ability to iteratively improve performance across 75 ML competitions. We noticed that MLE-Bench Lite is a carefully selected subset from MLE-Bench, providing an option that only requires about $\frac{1}{4}$ of the testing resources for teams with limited experimental resources while ensuring task diversity as much as possible. Meanwhile, the baseline methods referenced in this article were compared on MLE-Bench Lite. To ensure a fair comparison, we rigorously tested CLEAR on MLE-Bench Lite under the same conditions as the baseline method. Finally, We use MLE-Bench's Lite subset (22 tasks) as a lower-cost alternative to full evaluation.
\end{itemize}

\noindent The above two complementary datasets effectively evaluate the efficiency and cost of agent execution and optimization in ML research engineering during long-duration complex tasks.

\heading{Evaluation metrics}
For PaperBench, its task score is automatically evaluated based on a large model, which is the weighted mean of 8,316 leaf scoring nodes. 
For MLE-Bench Lite, we evaluated the dataset on 22 tasks, calculating the percentage of tasks that resulted in valid submissions out of the total number of tasks, the percentage of tasks that scored above the median, the percentage of tasks achieving bronze, silver and gold medals, and the percentage of tasks obtaining any medal.

\heading{Baselines}
We compare CLEAR against the following baselines on each dataset: 
\begin{itemize}[leftmargin=*]
\item On PaperBench, we compared our method with three different agent methods, namely AiScientist, BasicAgent, and IterAgent, under the same evaluation protocol, despite their distinct processes and structures.

\item On MLE-Bench Lite, we report comparative results against strong autonomous ML engineering systems, including methods such as AIDE \citep{jiang2025aide}, ML Master 2.0 \citep{zhu2026toward}, and LoongFlow \citep{wan2025loongflow}.
\end{itemize}

\heading{Implementation details}
Following the settings in AiScientist, we use Gemini-3-Flash and GLM-5 \citep{zeng2026glm} as the two backbone LLMs to instantiate the aforementioned baseline methods.
For our own method, CLEAR, to better meet the working needs of our designed model collaboration, we have chosen Claude-Opus 4.6, which possesses strong thinking capabilities, as the planning model $M_p$, while Claude-Sonnet 4.6, which has similar habits but is more cost-effective, has been selected as the coding model $M_c$.

The proposed visual lifecycle harness in the CLEAR features a human-in-the-loop checkpoint that allows users to inspect and revise specifications before code generation. We declare that: in the following experiments, CLEAR did not use manual review of system generated documents in all tests and did not make any modifications to the system generated content. 

In PaperBench and MLE-Bench Lite, each run is allocated an Nvidia A800 80GB GPU and a 24-hour budget for each task. For the comprehensive evaluation of PaperBench, we adopted the official evaluation protocol \citep{starace2025paperbench} and used GPT-5.4 as our judging model. All experimental data provided in this article are based on a fixed evaluation protocol, which ensures a fair comparison between different methods.

\begin{table*}[t]
\caption{Main results (\%) on MLE-Bench Lite. Bold and underlined denote the best and second-best results, respectively. The blue shaded line highlights CLEAR (Ours).}
\centering
\small
\setlength\tabcolsep{3.0pt}

\definecolor{headergray}{RGB}{245,245,245}
\definecolor{lightgray}{RGB}{250,250,250}
\definecolor{ourcol}{RGB}{235,242,255}

\rowcolors{2}{lightgray}{white}

\begin{tabular*}{\textwidth}{@{\extracolsep{\fill}}l l c c c c c c}
\toprule

\rowcolor{headergray}
\textbf{Method} & \textbf{Model} &
\textbf{Valid} & \textbf{Above Median} &
\textbf{Bronze} & \textbf{Silver} &
\textbf{Gold} & \textbf{Any Medal} \\

\midrule

AIDE & Gemini & 77.27 & 54.55 & 4.55 & 9.09 & 31.82 & 45.45 \\
LoongFlow & Gemini & 77.27 & 77.27 & 12.12 & 25.76 & \underline{39.39} & \underline{77.27} \\
AiScientist & Gemini & 100.00 & \underline{86.36} & \textbf{18.18} & 31.82 & 31.82 & \textbf{81.82} \\

AIDE & GLM & 77.27 & 50.00 & 4.55 & 13.64 & 22.73 & 40.91 \\
ML-Master 2.0 & GLM & 100.00 & 81.82 & \textbf{18.18} & 13.64 & \underline{31.82} & 63.64 \\
AiScientist & GLM & 100.00 & \textbf{90.91} & 9.09 & \underline{31.82} & \textbf{40.91} & \textbf{81.82} \\

\midrule

\rowcolor{ourcol}
CLEAR (Ours) & Opus \& Sonnet &
100.00 & \underline{86.36} & \underline{13.64} & \textbf{36.37} & 27.28 & \underline{77.27} \\

\bottomrule
\end{tabular*}
\vspace{2mm}
\label{tab:mlebench}
\end{table*}

\section{Experimental Results}
We evaluate CLEAR from three perspectives:
\begin{itemize}[leftmargin=*]
\item \textbf{RQ1: Effectiveness.} Can CLEAR reliably perform long-horizon ML research engineering across both RAR and MLE settings?
\item \textbf{RQ2: Efficiency.} Does CLEAR achieve a favorable performance--cost trade-off compared with existing ML research agents?
\item \textbf{RQ3: Mechanism.} How do structured specifications, specification-conditioned generation and specification-conditioned repair contribute to CLEAR's effectiveness and efficiency?
\end{itemize}

To answer the above three questions, we first evaluate overall effectiveness and efficiency on PaperBench and MLE-Bench Lite in Sections~\ref{main_paperbench} and~\ref{main_mlebench}. We then analyze the contributions of the key design choices in Section~\ref{ablation}.

Finally, as a supplement to the above experimental results, a section of token cost statistics will be shown in Section~\ref{token consumption}.

\vspace{-2mm}
\subsection{Main Results on PaperBench}
\label{main_paperbench}
Table~\ref{tab:paperbench} summarizes the overall performance on PaperBench. CLEAR achieves an average score of 29.87 across 20 tasks, ranking second on 10 tasks, while maintaining an average execution cost of only \$7.91 per task. These results indicate that CLEAR achieves a competitive performance--cost trade-off compared with existing agent frameworks.
We further analyze the comparison from three perspectives.
\begin{enumerate*}
\item \emph{Structured collaboration improves over lightweight agents.}
Compared with the BasicAgent baseline, CLEAR improves the average score from 22.58 to 29.87 (+32.29\%) with only a \$3.01 increase in average cost per task. This comparison shows that the additional computation introduced by the CLEAR workflow is accompanied by a substantial performance improvement while retaining relatively low execution cost.

\item \emph{More iterations do not necessarily yield better efficiency.} Compared with IterAgent, CLEAR achieves a higher average score while reducing the average cost by 85.6\% (\$7.91 vs.\ \$54.90). This comparison suggests that increasing iterative interaction alone does not necessarily yield a favorable performance--cost trade-off, whereas CLEAR maintains competitive performance with substantially lower execution cost.

\item \emph{Adaptive resource allocation provides a favorable alternative to scaling a single strong model.}
Compared with AiScientist using the same model family, CLEAR reduces the average cost by 35.2\% while retaining 88.6\% of the performance. The result suggests that dynamically allocating a variable number of coding workers under strong planning guidance is more cost-effective than uniformly increasing model capacity.

\end{enumerate*}

At the task level, CLEAR exhibits substantial variation across different PaperBench instances. It achieves competitive results on \textit{all-in-one}, \textit{bridging-data-gaps}, \textit{mechanistic-understanding}, and \textit{pinn}, where it ranks among the top-performing configurations. Meanwhile, larger performance gaps remain on tasks such as \textit{adaptive-pruning} and \textit{robust-clip}. These results indicate that the effectiveness of CLEAR is task-dependent and leave room for more adaptive task-specific generation and repair strategies.

\begin{figure}[t]
\centering
\includegraphics[width=\columnwidth]{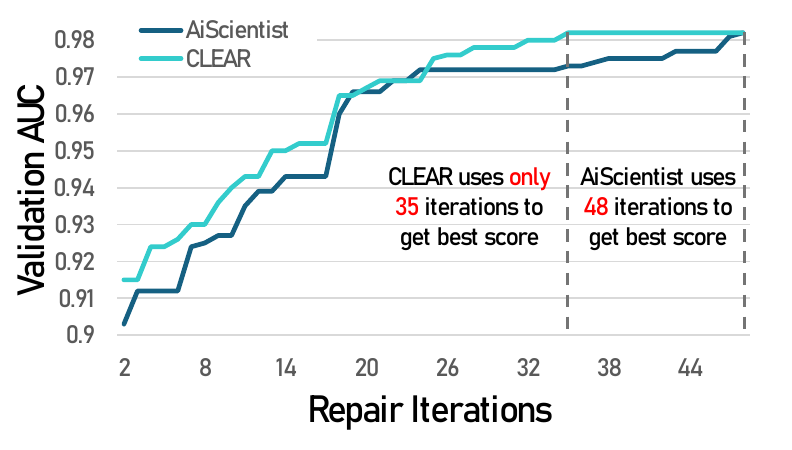}
\caption{CLEAR autonomously improving performance on a competition-style ML task over iterations. On MLE-Bench Lite’s Detecting Insults task, it completes 35 experiment cycles, reaching a validation AUC of 0.982 with 17 best-so-far updates. Compared to AiScientist, CLEAR reduces iterations by 27.08\%.}
\label{fig:iteration}

\end{figure}

\begin{table*}[t]
\centering
\caption{Repair dynamics of CLEAR on representative MLE-Bench Lite tasks. We report the number of repair iterations, best-score updates, and the proportion of useful repairs.}
\label{tab:paperbench_repair}
\small
\setlength{\tabcolsep}{6pt}

\definecolor{headergray}{RGB}{245,245,245}
\definecolor{lightgray}{RGB}{250,250,250}
\definecolor{ourcol}{RGB}{235,242,255}

\rowcolors{2}{lightgray}{white}

\begin{tabular}{l l r r r}
\toprule
\rowcolor{headergray}
\textbf{Task} &
\textbf{Task Type} &
\textbf{Repairs} &
\textbf{Best-score Updates} &
\textbf{Useful Repair Rate} \\
\midrule

aerial-cactus-identification
& Image Classification
& 37 & 13 & 35.14\% \\

detecting-insults-in-social-commentary
& Text Classification
& 35 & 17 & 48.57\% \\

nomad2018-predict-transparent-conductors
& Tabular
& 49 & 35 & \textbf{71.43\%} \\

text-normalization-challenge-english-language
& Seq2Seq
& 46 & 25 & 54.35\% \\

\midrule
\rowcolor{headergray}
\textbf{Average}
& --
& 41.75
& 22.50
& 52.37\% \\

\bottomrule
\end{tabular}
\vspace{2mm}
\end{table*}

\begin{table}[t]
\caption{Ablation results on PaperBench full evaluation. "CLEAR w/o specs" means CLEAR only uses $M_p$ to parse input documents, without processing by our designed skills. Bold and underlined denote the best and second-best results on each task.}
\centering
\small
\setlength\tabcolsep{2.5pt}
\definecolor{headergray}{RGB}{245,245,245}
\definecolor{lightgray}{RGB}{250,250,250}
\definecolor{ourcol}{RGB}{235,242,255}

\rowcolors{2}{lightgray}{white}

\begin{tabular*}{\columnwidth}{@{\extracolsep{\fill}}l l c c}
\toprule

\rowcolor{headergray}
\textbf{Method} & \textbf{Model} &
\textbf{Avg Score} & \textbf{Avg Cost} \\

\midrule

\rowcolor{ourcol}
CLEAR & Opus \& Sonnet & \underline{29.87} & \underline{\$7.91} \\
CLEAR w/o specs & Opus \& Sonnet & 24.14 & \$10.08 \\
CLEAR & GLM & 26.41 & \$11.95 \\
\bottomrule
\end{tabular*}
\label{tab:ablation_paperbench}
\end{table}

\vspace{-2mm}
\subsection{Main Results on MLE-Bench Lite}
\label{main_mlebench}
Table \ref{tab:mlebench} reports the main results of MLE-Bench Lite. Compared with the baseline method, CLEAR matches the best Valid Submission rate and remains competitive on Above Median and Any Medal. All 22 tasks formed valid submissions, with 86.36\% of tasks achieving scores above the median and 77.27\% of tasks winning medals. The above results support the effectiveness of CLEAR in executing ML research engineering tasks under MLE setting.

In terms of performance--cost trade-off, Figure~\ref{fig:iteration} compares the validation AUC trajectories of CLEAR and AiScientist on the Detecting Insults task from MLE-Bench Lite. Without human intervention, CLEAR reaches a validation AUC of 0.982 after 35 repair iterations, with 17 best-score updates. In comparison, AiScientist requires 48 repair iterations to reach the same level.
Specifically, the figure shows two main trends:
\begin{enumerate*}
\item CLEAR reaches the same validation AUC earlier in most stages, suggesting that the structured specification layer provides more effective guidance for downstream generation and repair;
\item AiScientist tends to make larger but sparser improvements, while CLEAR makes smaller and more frequent updates. This suggests that CLEAR's specification-conditioned repair process enables more fine-grained engineering optimization with fewer repair iterations.
\end{enumerate*}

Together with the PaperBench results, these findings answer \textbf{RQ1} and \textbf{RQ2}: CLEAR is effective across both RAR and MLE settings, and it achieves a favorable performance--cost trade-off by maintaining competitive task performance while reducing generation cost or repair iterations.

\vspace{-2mm}
\subsection{Mechanism Analysis}
\label{ablation}
\begin{figure}[t]
\centering
\includegraphics[width=\columnwidth]{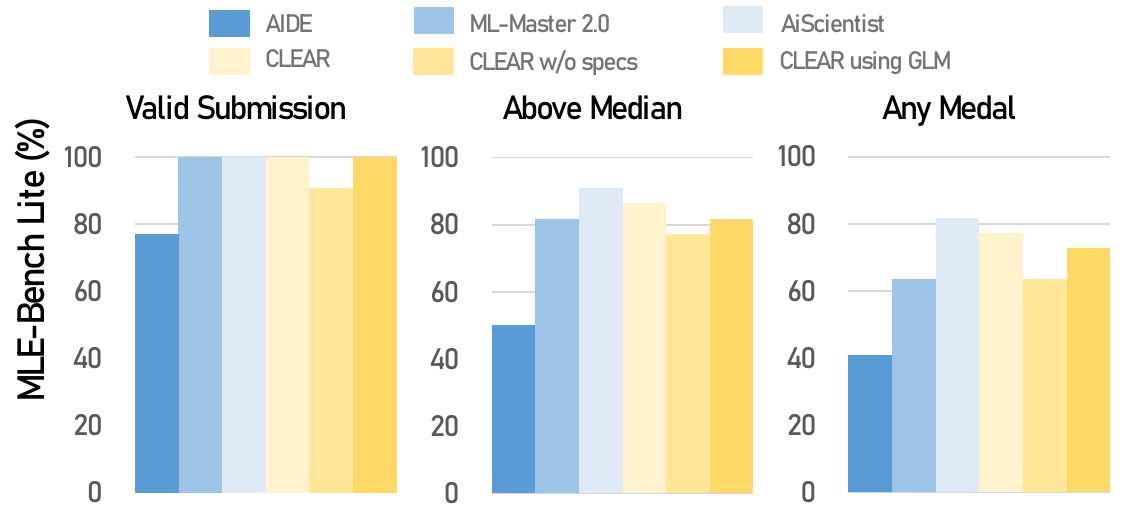}
\caption{Ablation results on MLE-Bench Lite. All values are percentages (\%). "CLEAR w/o specs" means CLEAR only uses $M_p$ to parse input documents, without processing by our designed skills. "CLEAR using GLM" means CLEAR uses GLM-5 for $M_p$ and $M_c$ at the same time.}
\label{fig:ablation_mlebench}
\vspace{-2mm}
\end{figure}

To answer \textbf{RQ3}, we examine the three key design choices of CLEAR: \emph{Document-to-specification parsing}, \emph{Specification-conditioned generation} and \emph{Specification-conditioned repair}.
Table~\ref{tab:ablation_paperbench} and Figure~\ref{fig:ablation_mlebench} report the corresponding results on PaperBench and MLE-Bench Lite.

\heading{Document-to-specification parsing}
We first remove the document-to-specification parsing skills, denoted as \emph{CLEAR w/o specs}, and let the planning model $M_p$ directly interpret the input documents. Removing structured specifications consistently degrades both effectiveness and efficiency. On PaperBench, the average score drops by 19.18\% while the average cost increases by 27.43\%. On MLE-Bench Lite, performance decreases by 9.09\%, 10.53\%, and 17.64\% on Valid Submission, Above Median, and Any Medal, respectively. These results support the role of $(P',S')$ as an explicit intermediate representation: turning the understanding of the task into more structured representation reduces ambiguity before implementation and avoids repeated interpretation during downstream generation and repair.

\heading{Specification-conditioned generation}
We next examine heterogeneous model assignment using \emph{CLEAR w/ GLM}, where both the planning model $M_p$ and coding model $M_c$ are replaced by GLM-5 while keeping the rest of the CLEAR architecture unchanged. This variant removes the heterogeneous model assignment while retaining the specification-centered workflow. This homogeneous GLM-5 configuration retains the specification-centered workflow but differs from the full CLEAR configuration in both model assignment and backbone. Compared with full CLEAR, its PaperBench score decreases by 11.59\% and its average cost increases by 51.07\%. On MLE-Bench Lite, it also underperforms CLEAR on Above Median and Any Medal, while achieving the same Valid Submission rate. These results show that the full Opus--Sonnet configuration achieves a better performance--cost trade-off than the homogeneous GLM-5 configuration.

Overall, the ablation results support both design principles of CLEAR: structured specifications improve the reliability and efficiency of task understanding, while heterogeneous model assignment improves the performance--cost trade-off by matching model capacity to stage-specific requirements.

\heading{Specification-conditioned repair}
\label{repair}
To better understand how CLEAR behaves during iterative repair, we analyze repair behavior on four representative MLE-Bench Lite tasks covering from image classification, text classification, tabular prediction to sequence-to-sequence learning.

As summarized in Table~\ref{tab:paperbench_repair}, CLEAR performs 35--49 repair iterations across these tasks, with an average of 41.75. The number of required iterations varies substantially across tasks, indicating that repair behavior will be different between tasks. 
We define a \emph{useful repair} as an iteration that improves the best validation score achieved so far. The useful-repair ratio ranges from 35.14\% to 71.43\%, averaging 52.37\%. Thus, more than half of the repair iterations on average improve the best-so-far validation score, although the frequency of such improvements varies across tasks. 
These results characterize CLEAR's behavior during iterative repair: across the analyzed tasks, a substantial fraction of bounded, localized repair iterations lead to measurable improvements within the predefined repair budget.

\begin{table}[t]
\caption{Token consumption of CLEAR and its two variants on PaperBench full evaluation.}
\centering
\small
\setlength\tabcolsep{2pt}

\definecolor{headergray}{RGB}{245,245,245}
\definecolor{lightgray}{RGB}{250,250,250}
\definecolor{ourcol}{RGB}{235,242,255}

\rowcolors{2}{lightgray}{white}

\begin{tabular}{l c c c}
\toprule

\rowcolor{headergray}
\textbf{Task Name} &
\multicolumn{3}{c}{\textbf{CLEAR}} \\
\cmidrule(lr){2-4}

&
\textbf{original} & \textbf{w/o specs} & \textbf{with GLM-5}\\
\midrule

adaptive-pruning & 517,006 & 622,068 & 3,022,296 \\
all-in-one & 647,319 & 963,726 & 2,727,358 \\
bam & 407,987 & 884,038 & 3,653,356 \\
bbox & 519,198 & 717,901 & 2,676,488 \\
bridging-data-gaps & 596,981 & 852,055 & 2,262,324 \\
fre & 690,222 & 1,065,629 & 4,050,093 \\
ftrl & 608,695 & 755,094 & 2,678,289 \\
lbcs & 497,164 & 850,478 & 2,726,949 \\
lca-on-the-line & 705,993 & 887,356 & 1,693,312 \\
mechanistic-understanding & 463,635 & 974,424 & 3,172,406 \\
pinn & 501,899 & 684,987 & 3,100,193 \\
rice & 706,867 & 1,021,489 & 3,116,092 \\
robust-clip & 551,805 & 699,040 & 1,684,801 \\
sample-specific-masks & 399,290 & 556,914 & 3,039,807 \\
sapg & 559,333 & 918,152 & 3,187,321 \\
sequential-neural & 389,731 & 1,062,345 & 3,678,553 \\
stay-on-topic & 596,446 & 768,587 & 2,913,926 \\
stochastic-interpolants & 768,990 & 988,850 & 4,610,972 \\
test-time-model-adaptation & 549,805 & 708,109 & 2,956,024 \\
what-will-my-model-forget & 608,380 & 851,134 & 2,805,173 \\

\midrule
\rowcolor{headergray}
\textbf{Average Token Cost} & 564,337 & 841,619 & 2,987,787\\

\bottomrule
\end{tabular}
\label{tab:paperbench_token}
\end{table}

\subsection{Token Consumption Analysis}
\label{token consumption}

Dollar cost provides a direct measure of practical deployment efficiency, but it depends on the pricing of the underlying models. To complement dollar cost with a pricing-independent usage indicator, we additionally report token consumption on the full PaperBench evaluation.

Table~\ref{tab:paperbench_token} compares CLEAR with its two variants. CLEAR consumes an average of 564K tokens per task, compared with 842K for CLEAR w/o specs and 2.99M for CLEAR w/ GLM-5. CLEAR w/o specs therefore consumes 49.1\% more tokens than the full CLEAR configuration. This observation is consistent with the hypothesis that maintaining an explicit intermediate specification reduces repeated interpretation during downstream generation and repair.
The homogeneous GLM-5 configuration consumes more than five times as many tokens as the full Opus--Sonnet configuration. This result is consistent with the overall efficiency advantage observed for the full CLEAR system. However, because the configurations use different model families and tokenizers, the cross-model token counts should be interpreted as a system-level usage comparison rather than a model-independent measure of computational cost.

\section{Related Work}
We review two lines of related work that underpin our study: LLM-based autonomous scientific research systems and multi-agent cooperation workflows for complex task execution.

\heading{Automated scientific research}
Recent work has explored the use of LLM agents for automating scientific research, including idea generation\citep{lu2025automated}, experiment execution\citep{huang2024mlagentbench}, code development\citep{zhang2025darwin,liDeepCodeOpenAgentic2025a,seo2026paper2codeautomatingcodegeneration}, and scientific writing\citep{wang2024autosurvey,lu2026towards}. These developments position autonomous scientific research as a realistic goal for general-purpose agent systems\citep{schmidgall2025agent,xu2026idea2story,tang2026ai}.

Among them, automating ML research engineering is particularly representative and challenging, as it requires transforming heterogeneous research artifacts into executable systems under limited cost and runtime constraints \citep{chan2025mle,zhu2026toward,chen2026toward}.
Recent systems such as BasicAgent~\citep{starace2025paperbench}, Leeroo~\citep{mohammadshahi2024leeroo}, AIDE~\citep{jiang2025aide}, AIBuildAI~\citep{zhang2026aibuildai}, and AiScientist~\citep{chen2026toward} demonstrate the feasibility of this paradigm. However, these systems often operate directly on heterogeneous research artifacts without an explicit intermediate representation, which can lead to cascading errors and expensive repair loops.

\heading{Multi-agent cooperation workflows}
Multi-agent cooperation has become a common paradigm for improving LLM agents on complex tasks. Early frameworks explore role-playing agents, dynamic team formation, and programmable multi-agent conversations~\citep{NEURIPS2023_a3621ee9,ICLR2024_578e65cd,wu2023autogenenablingnextgenllm}. These works show that task decomposition and agent specialization can improve problem solving, but coordination is often mediated through flexible natural-language interaction and relies on human-designed prompts heavily. 
In software engineering, multi-agent workflows have been adapted to code generation, repository editing, and issue resolution. MetaGPT and ChatDev organize agents through standardized software-development pipelines~\citep{hong2024metagpt,qian2024chatdevcommunicativeagentssoftware}. SWE-agent emphasizes agent-computer interfaces for repository-level editing and testing~\citep{NEURIPS2024_5a7c9475}, while AutoCodeRover and CodeR further incorporate code search, program analysis, and task graphs for issue resolution~\citep{10.1145/3650212.3680384,chen2024coderissueresolvingmultiagent}.

Despite recent advances, multi-agent systems still suffer from coordination failures, inter-agent misalignment, communication overhead, and weak verification~\citep{cemri2025multiagentllmsystemsfail,yan2025selftalkcommunicationcentricsurveyllmbased}. These issues become more severe in long-horizon ML research engineering, where task understanding, execution feedback, and repair decisions must remain consistent across stages. This motivates introducing a more explicit intermediate representation between research understanding and executable implementation.

\vspace{-2mm}
\section{Conclusion}
We presented CLEAR, a specification-centered framework for controllable long-horizon ML research engineering. CLEAR introduces an intermediate execution specification between heterogeneous research artifacts and executable implementations, enabling specification-conditioned planning, localized code generation, and targeted failure repair. Built upon this representation, CLEAR further integrates heterogeneous model assignment with lifecycle-oriented execution control, allowing different models to be dynamically allocated according to their capabilities and roles. Experiments on PaperBench and MLE-Bench Lite demonstrate that CLEAR achieves competitive performance while maintaining favorable performance--cost trade-offs across both research automation and machine learning engineering scenarios. Ablation studies further validate the importance of structured specifications and adaptive model collaboration. Overall, our results suggest that future ML research engineering agents can benefit from explicitly separating \emph{what should be implemented} from \emph{how it should be implemented, executed, and refined}.

\heading{Future Work}
Several directions remain open. First, the reliability of execution specifications could be further improved through automatic consistency checking, cross-modal verification, and execution-based validation. Second, the number of low-cost coding workers could be allocated more adaptively based on task difficulty, uncertainty, failure type, and expected repair value. Finally, the specification layer could be extended into a persistent research engineering memory that accumulates reusable implementation patterns, failure modes, and repair strategies across tasks.

\section{Acknowledgments}
This work was funded by the Strategic Priority Research Program of the CAS under Grants No. XDB0680102, the National Natural Science Foundation of China (NSFC) under Grants No. U25B2076, 62472408 and 62441229, and the National Key Research and Development Program of China under Grants No. 2023YFA1011602.
All content represents the opinion of the authors, which is not necessarily shared or endorsed by their respective employers and/or sponsors.

\section*{Ethical Considerations}
This article is aimed at the problem that in automatic ML research engineering, directly operating on heterogeneous research artifacts without an explicit intermediate representation causes cascading errors and expensive repair loops.
To mitigate this issue, we propose CLEAR, a controllable system for long-horizon ML research workflow in which a structured specification layer transforms heterogeneous research artifacts into human verifiable execution specifications. 
With respect to data usage, all experiments are conducted exclusively on publicly available academic benchmarks, including PaperBench and MLE-Bench Lite. 
No private, proprietary, or sensitive data is used at any stage of data collection, training, or evaluation. 
Furthermore, the proposed method does not introduce or expose any new personal information. 
As a result, this study does not raise concerns related to privacy, data misuse, or unauthorized access, and complies with common ethical standards for academic research in information retrieval and natural language processing.

\bibliographystyle{ACM-Reference-Format}
\bibliography{CLEAR}

@String{Computing = "Computing" }

@inproceedings{lu2025automated,
  title={Automated Capability Discovery via Model Self-Exploration},
  author={Lu, Cong and Hu, Shengran and Clune, Jeff},
  year={2025},
  booktitle={ICLR 2025 Workshop on Foundation Models in the Wild}
}

@inproceedings{huang2024mlagentbench,
  title={MLAgentBench: Evaluating Language Agents on Machine Learning Experimentation},
  author={Huang, Qian and Vora, Jian and Liang, Percy and Leskovec, Jure},
  booktitle={International Conference on Machine Learning},
  pages={20271--20309},
  year={2024},
  organization={PMLR}
}

@article{zhang2025darwin,
  title={Darwin godel machine: Open-ended evolution of self-improving agents},
  author={Zhang, Jenny and Hu, Shengran and Lu, Cong and Lange, Robert and Clune, Jeff},
  journal={arXiv preprint arXiv:2505.22954},
  year={2025}
}

@article{wang2024autosurvey,
  title={Autosurvey: Large language models can automatically write surveys},
  author={Wang, Yidong and Guo, Qi and Yao, Wenjin and Zhang, Hongbo and Zhang, Xin and Wu, Zhen and Zhang, Meishan and Dai, Xinyu and Zhang, Min and Wen, Qingsong and others},
  journal={Advances in neural information processing systems},
  volume={37},
  pages={115119--115145},
  year={2024}
}

@article{lu2026towards,
  title={Towards end-to-end automation of AI research},
  author={Lu, Chris and Lu, Cong and Lange, Robert Tjarko and Yamada, Yutaro and Hu, Shengran and Foerster, Jakob and Ha, David and Clune, Jeff},
  journal={Nature},
  volume={651},
  number={8107},
  pages={914--919},
  year={2026},
  publisher={Nature Publishing Group UK London}
}

@article{schmidgall2025agent,
  title={Agent laboratory: Using llm agents as research assistants},
  author={Schmidgall, Samuel and Su, Yusheng and Wang, Ze and Sun, Ximeng and Wu, Jialian and Yu, Xiaodong and Liu, Jiang and Moor, Michael and Liu, Zicheng and Barsoum, Emad},
  journal={Findings of the Association for Computational Linguistics: EMNLP 2025},
  pages={5977--6043},
  year={2025},
  publisher={Association for Computational Linguistics}
}

@article{xu2026idea2story,
  title={Idea2Story: An Automated Pipeline for Transforming Research Concepts into Complete Scientific Narratives},
  author={Xu, Tengyue and Qian, Zhuoyang and Liu, Gaoge and Ling, Li and Zhang, Zhentao and Wu, Biao and Zhang, Shuo and Lu, Ke and Shi, Wei and Wang, Ziqi and others},
  journal={arXiv preprint arXiv:2601.20833},
  year={2026}
}

@article{tang2026ai,
  title={Ai-researcher: Autonomous scientific innovation},
  author={Tang, Jiabin and Xia, Lianghao and Li, Zhonghang and Huang, Chao},
  journal={Advances in Neural Information Processing Systems},
  volume={38},
  pages={9481--9520},
  year={2026}
}

@inproceedings{chan2025mle,
  title={Mle-bench: Evaluating machine learning agents on machine learning engineering},
  author={Chan, Jun Shern and Chowdhury, Neil and Jaffe, Oliver and Aung, James and Sherburn, Dane and Mays, Evan and Starace, Giulio and Liu, Kevin and Maksin, Leon and Patwardhan, Tejal and others},
  booktitle={International Conference on Learning Representations},
  volume={2025},
  pages={50466--50494},
  year={2025}
}

@article{zhu2026toward,
  title={Toward ultra-long-horizon agentic science: Cognitive accumulation for machine learning engineering},
  author={Zhu, Xinyu and Cai, Yuzhu and Liu, Zexi and Zheng, Bingyang and Wang, Cheng and Ye, Rui and Chen, Jiaao and Wang, Hanrui and Wang, Wei-Chen and Zhang, Yuzhi and others},
  journal={arXiv preprint arXiv:2601.10402},
  year={2026}
}

@article{chen2026toward,
  title={Toward Autonomous Long-Horizon Engineering for ML Research},
  author={Chen, Guoxin and Chen, Jie and Chen, Lei and Zhao, Jiale and Meng, Fanzhe and Zhao, Wayne Xin and Song, Ruihua and Chen, Cheng and Wen, Ji-Rong and Jia, Kai},
  journal={arXiv preprint arXiv:2604.13018},
  year={2026}
}

@inproceedings{starace2025paperbench,
  title={PaperBench: Evaluating AI’s Ability to Replicate AI Research},
  author={Starace, Giulio and Jaffe, Oliver and Sherburn, Dane and Aung, James and Chan, Jun Shern and Maksin, Leon and Dias, Rachel and Mays, Evan and Kinsella, Benjamin and Thompson, Wyatt and others},
  booktitle={International Conference on Machine Learning},
  pages={56843--56873},
  year={2025},
  organization={PMLR}
}

@article{mohammadshahi2024leeroo,
  title={Leeroo orchestrator: Elevating llms performance through model integration},
  author={Mohammadshahi, Alireza and Shaikh, Ali and Yazdani, Majid},
  journal={arXiv preprint arXiv:2401.13979},
  volume={3},
  year={2024}
}

@article{jiang2025aide,
  title={Aide: Ai-driven exploration in the space of code},
  author={Jiang, Zhengyao and Schmidt, Dominik and Srikanth, Dhruv and Xu, Dixing and Kaplan, Ian and Jacenko, Deniss and Wu, Yuxiang},
  journal={arXiv preprint arXiv:2502.13138},
  year={2025}
}

@article{zhang2026aibuildai,
  title={AIBuildAI: An AI Agent for Automatically Building AI Models},
  author={Zhang, Ruiyi and Qin, Peijia and Cao, Qi and Zhang, Li and Xie, Pengtao},
  journal={arXiv preprint arXiv:2604.14455},
  year={2026}
}

@article{wan2025loongflow,
  title={Loongflow: Directed evolutionary search via a cognitive plan-execute-summarize paradigm},
  author={Wan, Chunhui and Dai, Xunan and Wang, Zhuo and Li, Minglei and Wang, Yanpeng and Mao, Yinan and Lan, Yu and Xiao, Zhiwen},
  journal={arXiv preprint arXiv:2512.24077},
  year={2025}
}

@article{zeng2026glm,
  title={Glm-5: from vibe coding to agentic engineering},
  author={Zeng, Aohan and Lv, Xin and Hou, Zhenyu and Du, Zhengxiao and Zheng, Qinkai and Chen, Bin and Yin, Da and Ge, Chendi and Huang, Chenghua and Xie, Chengxing and others},
  journal={arXiv preprint arXiv:2602.15763},
  year={2026}
}

@inproceedings{NEURIPS2023_a3621ee9,
  title = {{CAMEL}: Communicative Agents for {``Mind''} Exploration of Large Language Model Society},
  author = {Li, Guohao and Hammoud, Hasan and Itani, Hani and Khizbullin, Dmitrii and Ghanem, Bernard},
  booktitle = {Advances in Neural Information Processing Systems},
  editor = {Oh, A. and Naumann, T. and Globerson, A. and Saenko, K. and Hardt, M. and Levine, S.},
  year = {2023},
  volume = {36},
  pages = {51991--52008},
  publisher = {Curran Associates, Inc.},
  url = {https://proceedings.neurips.cc/paper_files/paper/2023/file/a3621ee907def47c1b952ade25c67698-Paper-Conference.pdf}
}

@inproceedings{ICLR2024_578e65cd,
  title = {{{AgentVerse}}: {{Facilitating}} Multi-Agent Collaboration and Exploring Emergent Behaviors},
  booktitle = {International Conference on Learning Representations},
  author = {Chen, Weize and Su, Yusheng and Zuo, Jingwei and Yang, Cheng and Yuan, Chenfei and Chan, Chi-Min and Yu, Heyang and Lu, Yaxi and Hung, Yi-Hsin and Qian, Chen and Qin, Yujia and Cong, Xin and Xie, Ruobing and Liu, Zhiyuan and Sun, Maosong and Zhou, Jie},
  editor = {Kim, B. and Yue, Y. and Chaudhuri, S. and Fragkiadaki, K. and Khan, M. and Sun, Y.},
  date = {2024},
  year = {2024},
  volume = {2024},
  pages = {20094--20136},
  url = {https://proceedings.iclr.cc/paper_files/paper/2024/file/578e65cdee35d00c708d4c64bce32971-Paper-Conference.pdf}
}

@misc{wu2023autogenenablingnextgenllm,
      title={AutoGen: Enabling Next-Gen LLM Applications via Multi-Agent Conversation}, 
      author={Qingyun Wu and Gagan Bansal and Jieyu Zhang and Yiran Wu and Beibin Li and Erkang Zhu and Li Jiang and Xiaoyun Zhang and Shaokun Zhang and Jiale Liu and Ahmed Hassan Awadallah and Ryen W White and Doug Burger and Chi Wang},
      year={2023},
      eprint={2308.08155},
      archivePrefix={arXiv},
      primaryClass={cs.AI},
      url={https://arxiv.org/abs/2308.08155}, 
}

@inproceedings{hong2024metagpt,
  title = {{MetaGPT}: Meta Programming for A Multi-Agent Collaborative Framework},
  author = {Hong, Sirui and Zhuge, Mingchen and Chen, Jiaqi and Zheng, Xiawu and Cheng, Yuheng and Zhang, Ceyao and Wang, Jinlin and Wang, Zili and Yau, Steven Ka Shing and Lin, Zijuan and Zhou, Liyang and Ran, Chenyu and Xiao, Lingfeng and Wu, Chenglin and Schmidhuber, J{\"u}rgen},
  booktitle = {International Conference on Learning Representations},
  year = {2024}
}

@misc{qian2024chatdevcommunicativeagentssoftware,
      title={ChatDev: Communicative Agents for Software Development}, 
      author={Chen Qian and Wei Liu and Hongzhang Liu and Nuo Chen and Yufan Dang and Jiahao Li and Cheng Yang and Weize Chen and Yusheng Su and Xin Cong and Juyuan Xu and Dahai Li and Zhiyuan Liu and Maosong Sun},
      year={2024},
      eprint={2307.07924},
      archivePrefix={arXiv},
      primaryClass={cs.SE},
      url={https://arxiv.org/abs/2307.07924}, 
}

@inproceedings{NEURIPS2024_5a7c9475,
  title = {{{SWE-agent}}: {{Agent-computer}} Interfaces Enable Automated Software Engineering},
  booktitle = {Advances in Neural Information Processing Systems},
  author = {Yang, John and Jimenez, Carlos and Wettig, Alexander and Lieret, Kilian and Yao, Shunyu and Narasimhan, Karthik and Press, Ofir},
  editor = {Globerson, A. and Mackey, L. and Belgrave, D. and Fan, A. and Paquet, U. and Tomczak, J. and Zhang, C.},
  date = {2024},
  year = {2024},
  volume = {37},
  pages = {50528--50652},
  publisher = {Curran Associates, Inc.},
  doi = {10.52202/079017-1601},
  url = {https://proceedings.neurips.cc/paper_files/paper/2024/file/5a7c947568c1b1328ccc5230172e1e7c-Paper-Conference.pdf}
}

@inproceedings{10.1145/3650212.3680384,
author = {Zhang, Yuntong and Ruan, Haifeng and Fan, Zhiyu and Roychoudhury, Abhik},
title = {AutoCodeRover: Autonomous Program Improvement},
year = {2024},
isbn = {9798400706127},
publisher = {Association for Computing Machinery},
address = {New York, NY, USA},
url = {https://doi.org/10.1145/3650212.3680384},
doi = {10.1145/3650212.3680384},
booktitle = {Proceedings of the 33rd ACM SIGSOFT International Symposium on Software Testing and Analysis},
pages = {1592–1604},
numpages = {13},
location = {Vienna, Austria},
series = {ISSTA 2024}
}

@misc{yan2025selftalkcommunicationcentricsurveyllmbased,
      title={Beyond Self-Talk: A Communication-Centric Survey of LLM-Based Multi-Agent Systems}, 
      author={Bingyu Yan and Zhibo Zhou and Litian Zhang and Lian Zhang and Ziyi Zhou and Dezhuang Miao and Zhoujun Li and Chaozhuo Li and Xiaoming Zhang},
      year={2025},
      eprint={2502.14321},
      archivePrefix={arXiv},
      primaryClass={cs.MA},
      url={https://arxiv.org/abs/2502.14321}, 
}

@misc{liDeepCodeOpenAgentic2025a,
  title = {{{DeepCode}}: {{Open Agentic Coding}}},
  shorttitle = {{{DeepCode}}},
  author = {Li, Zongwei and Li, Zhonghang and Guo, Zirui and Ren, Xubin and Huang, Chao},
  year = {2025},
  eprint = {2512.07921},
  eprinttype = {arXiv},
  eprintclass = {cs},
  doi = {10.48550/arXiv.2512.07921},
  url = {http://arxiv.org/abs/2512.07921},
  pubstate = {prepublished}
}

@misc{seo2026paper2codeautomatingcodegeneration,
      title={Paper2Code: Automating Code Generation from Scientific Papers in Machine Learning}, 
      author={Minju Seo and Jinheon Baek and Seongyun Lee and Sung Ju Hwang},
      year={2026},
      eprint={2504.17192},
      archivePrefix={arXiv},
      primaryClass={cs.CL},
      url={https://arxiv.org/abs/2504.17192}, 
}

@misc{chen2024coderissueresolvingmultiagent,
      title={CodeR: Issue Resolving with Multi-Agent and Task Graphs}, 
      author={Dong Chen and Shaoxin Lin and Muhan Zeng and Daoguang Zan and Jian-Gang Wang and Anton Cheshkov and Jun Sun and Hao Yu and Guoliang Dong and Artem Aliev and Jie Wang and Xiao Cheng and Guangtai Liang and Yuchi Ma and Pan Bian and Tao Xie and Qianxiang Wang},
      year={2024},
      eprint={2406.01304},
      archivePrefix={arXiv},
      primaryClass={cs.CL},
      url={https://arxiv.org/abs/2406.01304}, 
}

@misc{cemri2025multiagentllmsystemsfail,
      title={Why Do Multi-Agent LLM Systems Fail?}, 
      author={Mert Cemri and Melissa Z. Pan and Shuyi Yang and Lakshya A. Agrawal and Bhavya Chopra and Rishabh Tiwari and Kurt Keutzer and Aditya Parameswaran and Dan Klein and Kannan Ramchandran and Matei Zaharia and Joseph E. Gonzalez and Ion Stoica},
      year={2025},
      eprint={2503.13657},
      archivePrefix={arXiv},
      primaryClass={cs.AI},
      url={https://arxiv.org/abs/2503.13657}, 
}

\end{document}